\documentclass{aa}
\usepackage{epsfig,times}
\usepackage{amssymb}
\usepackage{txfonts}
\usepackage{graphicx}
\begin{document}

\title{Thermal radiation from magnetic neutron star surfaces}
\author{J.F. P\'erez--Azor\'{\i}n \and J.A.~Miralles 
\and J.A.~Pons }
%\offprints{}
\institute{Departament de F\'{\i}sica Aplicada, Universitat d'Alacant,
           Ap. Correus 99, 03080 Alacant, Spain}

\date{Received 7 July 2004/ Accepted 22 November 2004}

\abstract{We investigate the thermal emission from magnetic neutron star surfaces
in which the cohesive effects of the magnetic field have produced the condensation
of the atmosphere and the external layers. This may happen for sufficiently 
cool ($T\le 10^6$) atmospheres with moderately intense magnetic fields 
(about $10^{13}$ G for Fe atmospheres). The thermal emission from an isothermal
bare surface of a neutron star shows no remarkable spectral features, but
it is significantly depressed at energies below some threshold energy. However, since the thermal
conductivity is very different in the normal and parallel directions to the magnetic
field lines, the presence of the magnetic field is expected to produce a highly anisotropic
temperature distribution, depending on the magnetic field geometry. In this case,
the observed flux of such an object looks very similar to a BB spectrum, but depressed
by a nearly constant factor at all energies. This results in a systematic
underestimation of the area of the emitter (and therefore its size) by a factor 5-10 (2-3).

}

\titlerunning{Thermal radiation from magnetic NSs}
\authorrunning{P\'erez--Azor\'{\i}n, Miralles \& Pons}

\maketitle

%%%%%%%%%%%%%%%%%%%%%%%%%%%%%%%%%%%%%%%%%%%%%%%%%%%%%%%%%%%%%%%%%%%%%%%%
\section{Introduction}

Almost a decade after the discovery of the soft X-ray source RXJ185635-3754 
(\cite{WWN96}) using the ROSAT-PSPC, the thermal component associated with the
direct emission from a neutron star's surface has been detected in more than
20 X-ray sources. In many cases it is superimposed on a power--law tail,
but seven of these objects are well--characterized as simple blackbodies
with temperatures ranging between 60 and 100 eV. The apparent small
emitting surface of RXJ185635-3754, inferred from the best blackbody fit and
its parallax (\cite{WL02}) have led to speculation about its nature, and
whether its apparent smallness can be considered as evidence that the 
object is a strange star: a self--bound object made of up, down and strange quarks
(\cite{Pons02,Drake02}), or a standard, misaligned pulsar. Although
this new observational class (isolated compact stars) is probably the first real 
opportunity to place stringent constraints on the equation of state (EOS) of dense
matter from astrophysical measurements (see \cite{LP2001} for a review),
one must be cautious before concluding that an apparently small
X-ray source is a quark star, because the X-ray spectrum is not the only
information available.  Using the Hubble Space Telescope,
Walter \& Matthews (\cite{WM97}) subsequently identified an
optical source at 6060 \AA$\,$ and 3000 \AA, with a brightness only
about 7 times brighter than an extrapolation of a 62 eV X-ray
blackbody into the optical V band.
The observed optical fluxes have been confirmed by subsequent
observations from the 2-meter NTT (\cite{Neu98}) and new HST 
measurements (\cite{Pons02}).
Remarkably, the other three isolated compact X-ray sources
that have been detected in the optical band
(RX J0720.4-3125, RX J1308.6+2127, and RX J1605.3+3249)
also have a significant optical excess over the extrapolation
of the X-ray blackbody (a factor 5 to 14). None of them have yet been
detected as 1.4 GHz radio sources. Thus, 
RXJ185635-3754 is not an uncommon object, for it shares the same general
observational properties of other isolated neutron stars (blackbody
spectrum in X-ray, no evident spectral features, optical excess), except
for the fact that four of them have fairly long periods (8-22 s),
while RXJ185635-3754 is not variable, with a reported upper limit
on the pulse amplitude of $<1.3\%$ (\cite{Bur03}).

Since blackbodies are no more than a simple approximation of the true
emission mechanism, a step forward in understanding the thermal emission of
neutron stars, consists of computing model atmospheres for low magnetic
fields ($< 10^{11} G$) and emergent spectra for several compositions, masses 
and radius (\cite{Rom87,Mil92,RR96}).
The parameters that determine the shape of the observed spectrum are
the atmospheric composition, the red-shifted temperature, $T_\infty=
T_{\rm eff}/(1+z)$, the redshift $z$, the interstellar medium
absorption, $n_H$, and the ratio $R_\infty/d$, where $d$ is the
distance to the object and $R_\infty=R(1+z)$, $R$ being the radius of 
the star.  For a Planck spectrum the redshift
factors contribute only to an overall scale factor, so that the
redshift does not affect the models. For  realistic atmospheres
with spectral features, however, the redshift is an important parameter
because the identification of spectral lines would determine the $M/R$ ratio
and establish relevant constraints on the EOS of dense matter.
After a detailed investigation of the different parameters
that affect the observed spectra, Pons et al. (2002) concluded that
X-ray data alone do not allow us to establish severe constraints,
given the large number of degrees of freedom. The combination
of the X-ray spectra with the optical fluxes turned out to be much
more restrictive, as was suggested earlier by \cite{Pav96}.
The broadband spectral energy distribution (SED) of isothermal heavy 
element atmospheres can fit the observed
SED, but these models predict detectable absorption lines and edges.
It was hoped that the more sensitive spectrographs on Chandra and XMM--Newton
would be able to detect spectral features and help to understand the
nature of this objects, but no lines or absorption edges have been
seen in long exposures (\cite{Bur01,Drake02}). Only the presence of
broadband departures from a single blackbody spectrum has recently been
suggested (\cite{Bur03}).

The puzzle, then, is as follows: single temperature models based on heavy element
atmospheres explain the broadband spectrum and the lack of variability of the X-ray
spectrum, but they are in contradiction with the absence of spectral features.
Single blackbody models, besides being unrealistic, cannot
account for the systematic optical excess that is observed in 4 sources. An anisotropic
surface temperature distribution (the simplest case would be two blackbody components) can reconcile
X-ray and optical observations, because the quantity $(R_\infty/D)^2$ must
be interpreted as the solid angle subtended by the star's surface
area which is visible at some distance $D$. Thus the true value
of $R_\infty$ will generally be larger if the assumption that the
temperature is uniform on the surface is relaxed.
However, this is barely consistent with the lack of pulsation
of RXJ185635-3754, unless its magnetic and rotation axes are oriented
in a very particular way.

Nevertheless, there is an important ingredient missing in the above discussion,
the likely existence of strong magnetic fields.
The strong (10$^{12-13}$G) magnetic fields of pulsars
suggest that most, if not all, neutron stars should have similarly strong
fields. However, detailed atmosphere models with strong magnetic
fields are only available for hydrogen (\cite{Zav95}), because
reliable opacities and EOS have not yet been developed for heavier
elements and because of the complexity of modeling magnetic atmospheres with
arbitrary magnetic field structures.  For heavy-element dominated atmospheres, 
only approximate
treatments of magnetic Fe atmospheres exist (\cite{RRM97}), and the
results show that the spectra are globally much closer to a blackbody
than for light element atmospheres.  There is a case to be made
that RX~J185635-3754 is magnetized. The deep VLT image of the
target released by van Kerkwijk \& Kulkarni (\cite{KK00}) shows what
looks like a classic bow-shock nebula. The presence of a bow-shock suggests
that this is a magnetized neutron star with a relativistic wind, as seen in the pulsars
PSR 1957+20 (Kulkarni \& Hester \cite{KH88}) and PSR 2224+65
(Cordes, Romani, \& Lundgren \cite{CRL93}).

An alternative possibility that explains naturally the absence of spectral
features is the emission from a solid surface.
This was a common idea 20-30 years ago (\cite{Bri80}, B80 in the following) until the 
existence of a thin gaseous atmosphere was appreciated and model atmospheres became more popular.
However, at sufficiently low temperatures, highly magnetized neutron stars may undergo a phase transition that
turns the gaseous atmosphere into a solid (\cite{Lai01}). The critical temperature below
which the atmosphere condensates depends on the composition and the magnetic field. For
example, for typical magnetic field strengths of $10^{13}$ G, a Fe atmosphere will
condensate for $T< 0.1$ keV while a H atmosphere needs temperatures lower than
0.03 keV to undergo the phase transition to the {\it metallic} state (\cite{Lai01}).
Notice that effective temperatures of the observed isolated neutron stars fall in this
temperature range, therefore they should plausibly be in the solid state if the dominant 
element in the atmosphere is Fe.
In such a metallic neutron star surface made of nuclei with atomic number $Z$
and atomic weight $A$, the pressure vanishes at a finite density
\begin{equation}
\rho_s \approx 560~A Z^{-3/5} B_{12}^{6/5} {\rm g\, cm}^{-3}
\label{rhos}
\end{equation}
where $B_{12}$ is the magnetic field in units of $10^{12}$ G.

This idea has been
recently revisited by Turolla et al. (2004), who found that the emitted
spectrum is strongly depressed at low energies, thus making more difficult the
reconciliation between observational data and the condensed surface model.
However, in this last work the emissivity is calculated neglecting one of the
transmission modes in some cases. This simplification is inaccurate and can modify
the emitted spectrum, as we will discuss in the text. 
Besides, the sole presence of a strong magnetic field, necessary 
to condensate the atmosphere, results naturally in an anisotropic surface temperature
(\cite{GKP04}), which must be calculated consistently with the magnetic field structure.
In this paper our aim is to study the thermal emission from solid surfaces of neutron stars
and its implications on the observational properties of isolated compact objects.

The paper is organized as follows: Sect. 2 is devoted to discussing the calculation of the emissivity 
starting from the description of the dielectric tensor for the condensed neutron star surface.
In Sect. 3 we show results, including integrated spectra for
different strengths and geometries of the magnetic field.
Final remarks and a summary of the main conclusions are discussed in Sect. 4.

%%%%%%%%%%%%%%%%%%%%%%%%%%%%%%%%%%%%%%%%%%%%%%%%%%%%%%%%%%%%%%%%%%%%%%%%
\section{Equations and input microphysics}

\subsection{Dielectric Tensor}

The emission properties (reflectivity, emissivity) of a
neutron star surface in a condensed state are obtained from the 
dielectric tensor, and were first
studied in detail in B80. In the presence of the
strong magnetic fields expected to exist at the surface of a
neutron star, the dielectric tensor changes significantly with respect to the
non-magnetic case, leading to birefringence and other associated phenomena.
In order to calculate the dielectric tensor in the presence of magnetic fields and 
to take into account dissipative processes, it is better to obtain
first the conductivity tensor $\sigma_{ij}$, and then to calculate
the dielectric tensor $\epsilon_{ij}$ according to the equation
\begin{equation}
\epsilon_{ij} = \delta_{ij}+i
\frac{4\pi}{\omega}\sigma_{ij}.
\end{equation}
In this expression, $\omega$ is the angular frequency of the electromagnetic
wave propagating in the medium.

For a magnetized plasma, electrical conductivities for the static case ($\omega=0$)
have been calculated by some authors (\cite{Her84,Pot99}) with and without taking 
into account the quantizing effect of the magnetic field. The general expression 
for the conductivity tensor can be written as follows
\begin{equation}
\sigma_{ij}=\int e^2 \frac{{\cal N}_B}{\epsilon/c^2}
\tau_{ij}\left(-\frac{\partial f_0}{\partial \epsilon}\right) d\epsilon,
\end{equation}
where $\epsilon$ is the electron energy, $f_0$ the Fermi-Dirac distribution, ${\cal N}_B$
is given by the expression (see \cite{Pot99} for details)
\begin{equation}
{\cal N}_B=\frac{m_e\omega_B}{2(\pi\hbar)^2}\sum_{n=0}^{n_{\rm max}}g_n
\sqrt{(\epsilon/c)^2-(m_ec)^2-2m_e\hbar\omega_Bn},
\end{equation}
and the functions $\tau_{ij}$ play the role of effective relaxation times.
For a non-quantizing magnetic field, these effective relaxation times 
can be expressed in terms of a relaxation time, $\tau_0$, which is the inverse 
of the collisional frequency $\nu^D$ (D after damping)
However, in the quantizing case, 
collisional frequencies in the longitudinal $\nu^D_{\parallel}$ and perpendicular 
$\nu^D_{\perp}$ direction to the magnetic field are no longer equal. 
For a quantizing magnetic field, the components of the $\tau_{ij}$
tensor are 
\begin{eqnarray}
\tau_{zz} = \tau_{\parallel}; \;\;\tau_{xx}=\frac{\tau_{\perp}}{1+(\omega_{B}
\tau_{\perp})^{2}}\; ; \; \; \tau_{yx}=\frac{\omega_{B}\tau_{\perp}^{2}}
{1+(\omega_{B}\tau_{\perp})^{2}}
\end{eqnarray}
with $\tau_{\parallel}$ and $\tau_{\perp}$ being the inverse of the
effective collisional frequencies, $\nu^D_{\parallel}$ and
$\nu^D_{\perp}$, respectively, and $\omega_B = \frac{eB}{m_e c}$ being
the electron cyclotron frequency.

In the dynamic case ($\omega\neq 0$), the conductivity tensor can be obtained by
the transformation
\begin{eqnarray}
%\tau_{\perp} \rightarrow \frac{\tau_{\perp}}{1-i\omega \tau_{\perp}} \\
%\tau_{\parallel} \rightarrow \frac{\tau_{\parallel}}{1-i\omega \tau_{\parallel}},
\tau_{\perp}^{-1} \rightarrow \tau_{\perp}^{-1}-i\omega \\
\tau_{\parallel}^{-1} \rightarrow \tau_{\parallel}^{-1}-i\omega
\end{eqnarray}
and the dielectric tensor, for degenerate non-relativistic electrons, 
reads

\begin{eqnarray}
\label{eqn:dielectric}
\epsilon_{ij} = \delta_{ij}+i
\frac{4\pi}{\omega}\sigma_{ij} =
            \left( \begin{array}{ccc}
                     S & -iD  & 0 \\
                    iD &   S  & 0 \\
                     0 &   0  & P
             \end{array} \right),
\end{eqnarray}
where we have defined the following complex quantities
\begin{eqnarray}
\left( \begin{array}{c} R \\ L \end{array} \right) &=& 1 - \frac{\omega_{p}^{2}}{\omega^{2}}
\frac{\omega}{\omega \mp \omega_{B}+i\nu^D_{\perp}}; \\
P &=& 1- \frac{\omega_{p}^{2}}{\omega^{2}+i\omega \nu^D_{\parallel}}; \\
S &=& \frac{1}{2}(R+L); \\ D &=& \frac{1}{2}(R-L).
\end{eqnarray}
Here, $\omega_p = (4 \pi e^2 n_e /m_e)^{1/2}$ is the electron plasma frequency
and, using the density at zero pressure given by Eq. (\ref{rhos}), the
electron particle density $n_e$ can be calculated as follows
\begin{equation}
n_e = 1.24 \times 10^{27} Z_{26}^{2/5} B_{12}^{6/5} {\rm cm}^{-3}.
\end{equation}

%%%%%%%%%%%%%%%%%%%%%%%%% FIGURE %%%%%%%%%%%%%%%%%%%%%%%%%%%%%%%%%
\begin{figure*}
\resizebox{\hsize}{!}{\includegraphics{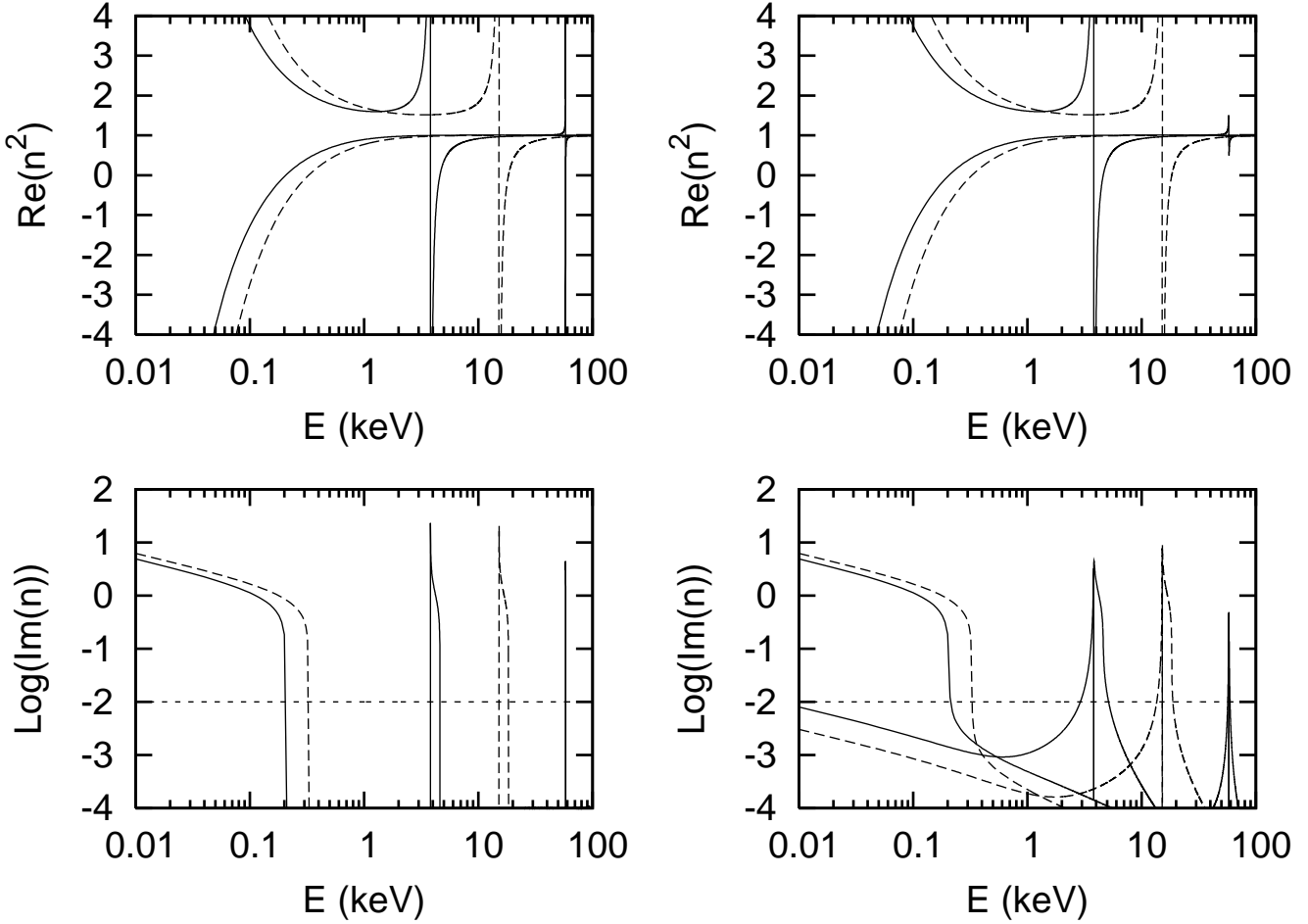}}
\caption{
Refractive indexes without (left) and with (right) damping effects.
The real parts are shown in the top panels and the imaginary parts
are shown at the bottom. Two different values of the magnetic field
are depicted, $5 \times 10^{12}$ G(solid lines) and $5 \times 10^{13}$ G
(dashed lines). The photon incidence angle has been taken to 43$^\circ$
to compare with Fig. 6 of Turolla et al. \cite{TZD04}.}
\label{indice}
\end{figure*}
%%%%%%%%%%%%%%%%%%%%%%%%% END FIGURE %%%%%%%%%%%%%%%%%%%%%%%%%%%%%%%%%

%%%%%%%%%%%%%%%%%%%%%%%%%%%%%%%%%%
\subsection{The dispersion relation.}

Consider an electromagnetic wave propagating in a medium described by a
dielectric tensor $\epsilon_{\alpha\beta}$.
The dispersion relation is easily obtained by introducing the
Maxwell tensor
\begin{equation}
\Lambda_{ij} = k_{i}k_{j}-k^{2}\delta_{ij}+\frac{\omega^{2}}{c^{2}}\epsilon_{ij},
\end{equation}
where $k_{i}$ are the Cartesian components of the wave vector.
Since, in terms of the Maxwell or dispersion tensor, the electric field of the wave satisfies
the equation
\begin{equation}
\Lambda_{ij}E_{j} = 0,
\label{max1}
\end{equation}
the condition to have a non-trivial solution leads to the dispersion equation
\begin{equation}
\mid \Lambda_{ij}\mid=0.
\end{equation}

In the same Cartesian frame as used in B80, where the magnetic field ${\bf B}$
is in the $x-z$ plane and forms an angle $\alpha$ with the $z$-axis, which is normal to
the surface, the dielectric tensor, obtained by a rotation transformation from 
Eq. (\ref{eqn:dielectric}), adopts the form
\begin{eqnarray}
\label{eqn:dielectric2}
\epsilon_{ij} =
        \left( \begin{array}{ccc}
                S\cos^{2}\alpha+P\sin^{2}\alpha & -iD\cos \alpha & \sin \alpha \cos \alpha (P-S) \\
                iD\cos\alpha  & S & -iD\sin \alpha  \\
                \sin \alpha \cos \alpha (P-S) & iD\sin \alpha & P\cos^{2}\alpha+S\sin^{2}\alpha
        \end{array} \right)
\end{eqnarray}
For our purposes it is convenient to write the dispersion relation for the transmitted wave
in terms of the incident one. We consider an incident wave with wave vector
\begin{equation}
\vec{k}_i = k_{i} (-\sin i \cos \beta, \sin i \sin \beta, -\cos i),
\end{equation}
where $i$ is the angle of incidence and $\beta$ the azimuth.
Since the wave is propagating in vacuum and we
are neglecting vacuum polarization,  the dispersion relation gives $k_{i}=\omega/c$.
The reflected and transmitted waves have wave-vectors
\begin{eqnarray}
\vec{k}_r &=& k_{i} (-\sin i \cos \beta, \sin i \sin \beta, \cos i) \nonumber \\
\vec{k}_t &=& k_{t} (-\sin \theta_m \cos \beta, \sin \theta_m \sin \beta, -\cos \theta_m); \; \; \; m = 1,2,
\end{eqnarray}
where the subscript $m$ refers to the ordinary and extraordinary modes and 
$\theta_m$ is the angle of refraction, which is a complex number.
Introducing the complex refractive index for the transmitted wave $n=kc/\omega$, 
and using the Snell law, $\sin \theta_{m} = \sin i / n_{m}$, the dispersion relation
leads to a quartic equation for the refractive index $n$,
\begin{eqnarray}
n^{4}(P+v\sin^{2} \alpha )+n^{2}(gv - 2PS+u\sin^{2}\alpha)+PRL+gu= \nonumber \\
\sin i \sin(2\alpha)\cos\beta(n^{2}-\sin^{2}i)^{1/2}(u+n^{2}v)
\label{eq:cuartica}
\end{eqnarray}
where $v=S-P$,  $u=PS-RL$, and  $g=\sin^{2}i [1-\sin^{2}\alpha(1+\cos^{2}\beta)]$.
This quartic equation can be solved analytically, after squaring, and the refractive
indexes for both ordinary and extraordinary modes can be obtained after analyzing
which are the two physical roots that satisfy the original Eq. (\ref{eq:cuartica}).

In Fig. \ref{indice} we show the real and imaginary parts of the two modes 
as a function of the photon energy
and for two different magnetic fields $B=5 \times 10^{12}$ G (solid lines) and
$B=5 \times 10^{13}$ G (dashes). The corresponding electron plasma frequencies
are 3.6 keV and 14.3 keV, respectively. We have taken a magnetic field normal
to the emitting surface and an incidence angle of 43$^\circ$. 
The left panels correspond to the case without
damping ($\nu_\parallel^D=\nu_\perp^D=0$) while the right panels show the 
equivalent results including effects of collisional damping (these collisional
damping frequencies were calculated by using the public code developed by 
Potekhin \footnote{{\tt www.ioffe.rssi.ru/astro/conduct/condmag.html}}).
The resonance at the plasma frequency $\omega_p$
is clearly visible, as well as a second, narrow resonance at $\omega_B$
($\approx$ 58 keV for $B=5 \times 10^{12} G$).
More interestingly, we can observe a region at low energy in which one of the
modes takes a large imaginary part, which will result in significant absorption.
The energy at which this happens can be estimated as follows. For simplicity,
let us assume normal incidence and a magnetic field normal to the surface, and
let us consider the case without damping. 
In this limit the dispersion relation is 
$$
P n^4  - 2PS n^2 + PRL = 0,
$$
and the roots are simply $n^2=L~$ and $n^2=R~$, i.e.,
\begin{eqnarray}
n^2 =1-\frac{\omega_p^2}{\omega(\omega \pm \omega_B)}.
\label{n2app}
\end{eqnarray}
Therefore, for energies
$\omega < \frac{\omega_B}{2}\left[ -1+\sqrt{1+4(\omega_p/\omega_B)^2} \right]$
the second mode acquires 
an increasingly larger imaginary part. It turns out that for $B > 10^{12}$ G  we are 
always in the case 
$\omega_p\ll\omega_B$, and the above condition becomes approximately $\omega <\omega_p^2/\omega_B$. 
In brief, we can expect two main features in the spectrum, 
a resonant absorption near the plasma and cyclotron frequencies, and lower
emission (compared to the BB) below a certain threshold energy 
$\approx \omega_p^2/\omega_B$. In Turolla et al. (\cite{TZD04}), the modes with imaginary
part larger than a certain value (0.01) were neglected, this is indicated
in the figure by the horizontal dashed line. In some cases, this may result in 
lower emissivity at low frequencies, and we prefer not to neglect them.

%%%%%%%%%%%%%%%%%%%%%%%%%%%%%%%%%%%%%%%%%%%%%%%%%%%%%%%%%%%%%%%%%%
\subsection{Reflectivity and emissivity}

Knowing the complex refractive indices of the two modes, and
solving Eq. (\ref{max1}), the ratio between the electric field 
components, $(E^\prime_x,E^\prime_y,E^\prime_x)$, of the transmitted wave 
can be obtained:
\begin{eqnarray}
&\frac{E^{'}_{m,x}}{E^{'}_{m,z}} &\equiv a_{m} = 
\left( D^{2}\sin^{2}\alpha + S\sin^{2}i - n_{m}^{2}\sin^{2}i \cos^{2}\beta
\right .
\nonumber \\
&+& \left. (P\cos^{2}\alpha+S\sin^{2}\alpha)(n_{m}^{2}-S-\sin^{2}i \sin^{2}\beta) \right)
\times
\nonumber \\ 
&& \left( \left[\sin i \cos \beta \sqrt{n^{2}_{m}-\sin^{2} i} 
+ \sin \alpha \cos \alpha (P-S)\right] (S-n_{m}^{2}) \right.+ 
\nonumber \\ 
&+& \left . i D \sin i \sin \beta \left(\cos \alpha \sqrt{n^{2}_{m}-\sin^{2} i} 
+ \sin \alpha \sin i \cos \beta \right)+\right.
\nonumber \\ 
&+&  \left. \sin \alpha \cos \alpha \left[(P-S) \sin^{2}i \sin^{2}\beta + D^{2}\right] \right)^{-1}
\label{eq:amplx}
\end{eqnarray}
and
\begin{eqnarray}
\frac{E^{'}_{m,y}}{E^{'}_{m,z}} &\equiv& b_{m} = 
\left[ a_{m} (\sin^{2}i \cos \beta \sin \beta -iD\cos \alpha)
\right.
\nonumber \\
&+& \left.
\sin \beta \sin i \sqrt{n^{2}_{m}-\sin^{2}i}+iD\sin \alpha \right] \nonumber \\
&& \times \left( \sin^{2}\beta \sin^{2} i - n^{2}_{m}+S \right)^{-1}. 
\label{eq:amply}
\end{eqnarray}
These relative amplitudes coincide, after some minor algebraic manipulation, with those in B80. 
In Turolla et al. (\cite{TZD04}), a different linear combination of the three equations
arising from Eq. (\ref{max1}) was used, but both results can be shown to be equivalent
when the refraction index is a solution of Eq. (\ref{eq:cuartica}).

%%%%%%%%%%%%%%%%%%%%%%%%% FIGURE %%%%%%%%%%%%%%%%%%%%%%%%%%%%%%%%%
\begin{figure}
\resizebox{\hsize}{!}{\includegraphics{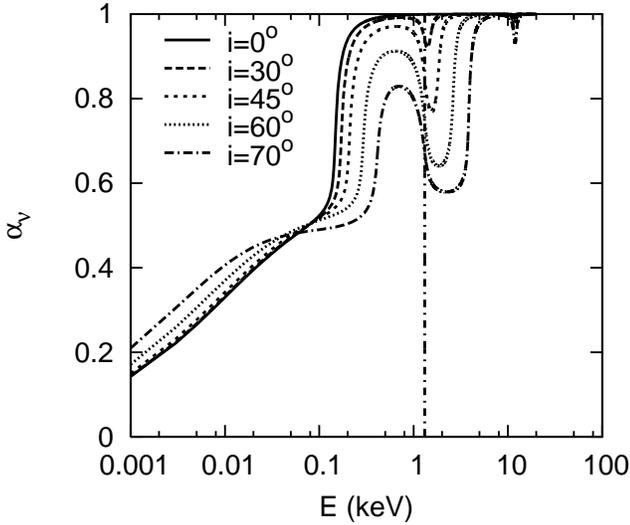}}
\caption{Emissivity normalized to the BB emission as a function
of energy for $B= 10^{12}$ G and $T=10^{6}$ K and varying the incident angle. 
The magnetic field has been taken to be normal to the emitting surface.
}
\label{figure2}
\end{figure}
%%%%%%%%%%%%%%%%%%%%%%%%% END FIGURE %%%%%%%%%%%%%%%%%%%%%%%%%%%%%%%%%

The boundary conditions at the surface of separation of both media,
(vacuum and the magnetized plasma) imply that the
tangential component of the magnetic and electric field and the normal component
of the magnetic and the electric displacement must be continuous. This results
in the following equation
\begin{eqnarray}
\left( \begin{array}{c} E_{\perp} \\ E_{\perp} \\ E_{\parallel} \\ E_{\parallel} \end{array} \right) =
\left( \begin{array}{cccc} 
B_1 & B_{2} & -1 & 0 \\
\frac{n_{1} \cos \theta_{1} B_1}{\cos i} & \frac{n_{2} \cos \theta_{2} B_2}{\cos i} & 1 & 0 \\ 
\frac{A_1}{\cos i}  & \frac{A_2}{\cos i} & 0 & 1 \\ 
\frac{C_1}{\sin i}  & \frac{C_2}{\sin i} & 0 & -1 \end{array}
\right) \cdot 
\left( \begin{array}{c} E^{'}_{1z} \\ E^{'}_{2z} \\ E^{''}_{\perp} \\ E^{''}_{\parallel} 
\end{array} \right)
\label{systemE}
\end{eqnarray}
where $E$, $E^\prime$ and $E^{\prime\prime}$ are, respectively, the electric
field of the incident, transmitted and reflected wave, and the subscripts
$\parallel$, $\perp$ refer to components parallel or perpendicular to the
incidence plane. Above, we have defined
$A_{m} = b_{m}\sin \beta -a_{m} \cos \beta$, 
$B_{m} = b_{m}\cos \beta +a_{m} \sin \beta$, and
$C_{m}=\epsilon_{31}a_{m}+\epsilon_{32}b_{m}+\epsilon_{33}$.
This system of equations can be solved for the electric field of the
reflected wave in terms of the incident one. Details and explicit 
expressions are given in the Appendix.

%%%%%%%%%%%%%%%%%%%%%%%%% FIGURE %%%%%%%%%%%%%%%%%%%%%%%%%%%%%%%%%
\begin{figure}
\resizebox{\hsize}{!}{\includegraphics{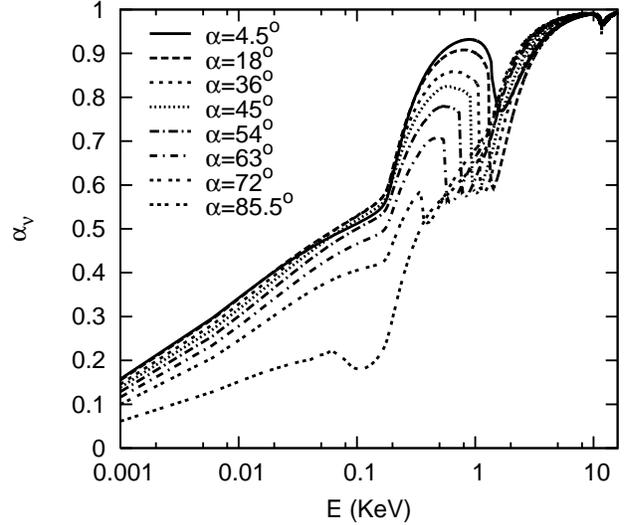}}
\caption{Normalized emissivity integrated over all possible incident angles as 
a function of energy for $B=10^{12}$ G and $T=10^{6}$ K. Different orientations
of the magnetic field are compared: from top to bottom 
$\theta = 4.5, 18, 36, 45, 54, 63, 72;$ and  $85.5^\circ$.
}
\label{spcf12}
\end{figure}
%%%%%%%%%%%%%%%%%%%%%%%%% END FIGURE %%%%%%%%%%%%%%%%%%%%%%%%%%%%%%%%%
The reflectivity can be now obtained by assuming that the incident wave is the
incoherent sum of two linearly polarized waves parallel and perpendicular
to the incidence plane. From Eq. (\ref{eq:campos}) and taking $E_\perp=1$,
$E_\parallel=0$, the two complex components of the reflected field
($E_{\parallel}^{''}$ and $E_{\perp}^{''}$) can be calculated, 
and the reflectivity is simply
\begin{eqnarray}
\rho_{\perp} = |E_{\parallel}^{''}|^{2} + |E_{\perp}^{''}|^{2}.
\end{eqnarray}
Analogously, for the case of polarization in the incidence plane, one takes $E_\perp=0$,
$E_\parallel=1$, and after obtaining $E_{\parallel}^{''}$ and $E_{\perp}^{''}$ 
the reflectivity is given by
\begin{eqnarray}
\rho_{\parallel} = |E_{\parallel}^{''}|^{2} + |E_{\perp}^{''}|^{2}.
\end{eqnarray}
To obtain the reflectivity of the unpolarized incident, $\rho_\nu$, wave we 
just take the average
\begin{eqnarray}
\rho_\nu = \frac{1}{2}(\rho_{\perp}+\rho_{\parallel})
\end{eqnarray}

Finally, the reflectivity is related to the emitted specific 
intensity $I_\nu$ by Kirchhoff's law
\begin{equation}
I_\nu=(1-\rho_\nu) B_\nu,
\end{equation}
where $B_\nu$ is the Planck intensity
\begin{equation}
B_\nu = \frac{2 h \nu^{3} /c^{2}}{\exp(h\nu /kT)-1}.
\end{equation}

In Fig. \ref{figure2} we illustrate how the normalized emissivity, i.e., 
emitted intensity normalized to the BB value, 
$\alpha_{\nu} \equiv I_{\nu}/B_{\nu}=1-\rho_\nu$,
varies with the angle
of incidence for a fixed magnetic field of $B=10^{12}$ G normal
to the surface. The temperature is $T=10^{6}$ K. The emissivity is strongly 
reduced compared to the blackbody case for energies lower than 0.2-0.5 keV, depending on 
the incident angle. A reduction for energies close to that corresponding to
the plasma frequency (indicated by the vertical line) is also evident in all cases.
Notice how for incident angles close to $\pi/2$ the emission is strongly
suppressed in a wide range of energies.

%%%%%%%%%%%%%%%%%%%%%%%%% FIGURE %%%%%%%%%%%%%%%%%%%%%%%%%%%%%%%%%
\begin{figure}
\resizebox{\hsize}{!}{\includegraphics{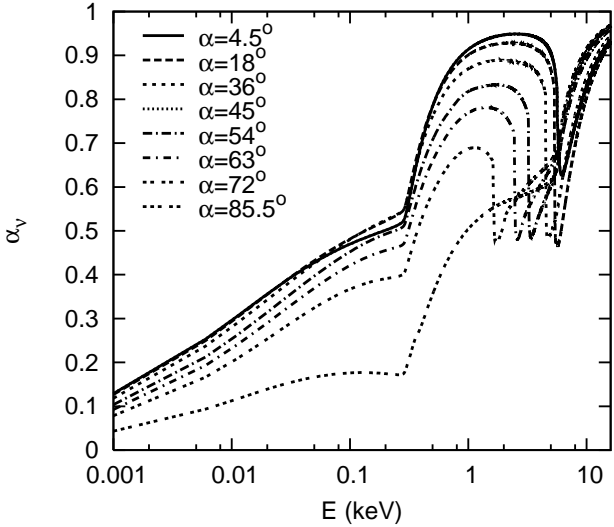}}
\caption{Same as Fig. \ref{spcf12} but for $B=10^{13}$ G.
}
\label{spcf13}
\end{figure}
%%%%%%%%%%%%%%%%%%%%%%%%% END FIGURE %%%%%%%%%%%%%%%%%%%%%%%%%%%%%%%%%

Let us now comment on the variation of the emissivity with other relevant
parameters.
In Fig. \ref{spcf12} we show the normalized emissivity, integrated over all
possible incident angles, as a function of the photon energy.
We have taken  $T=10^{6}$ K and $B=10^{12}$ G and we show results for 
different magnetic field orientations 
($\alpha$ is the angle between the magnetic field and a vector normal to the surface).
At energies greater than the electron plasma frequency (about 1 keV in this case) 
the emitted flux approaches the BB value ($\alpha_\nu=1$), but the spectrum is 
significantly depressed at low energies. As explained above, this is due to the fact
that the refractive index has a large imaginary part.
A resonance, which produces a reduction of the emissivity, is also visible at 
energies close to the electron plasma frequency.  
Notice that the resonant energy is not exactly the plasma frequency
but depends on the angle between the magnetic field and the surface. 
This can be understood considering that the refractive index of one of the
modes becomes infinity when the coefficient of $n^4$ in the dispersion
relation vanishes. Again, neglecting
damping for simplicity, we have $P + (S-P)\sin^2\alpha =0$, which leads to
\begin{equation}
\omega^4 - (\omega_B^2 + \omega_p^2) \omega^2 + \omega_p^2 \omega_B^2 \cos^2\alpha =0.
\end{equation}
The solution of this biquartic equation is
\begin{equation}
\omega_{\pm}^2 = \frac{\omega_B^2 + \omega_p^2}{2} \left[
1 \pm \left( 1 - \frac{4 \omega_p^2 \omega_B^2 \cos^2\alpha}
{(\omega_B^2 + \omega_p^2)^2} \right)^{1/2}
\right]
\end{equation}
For magnetic fields perpendicular to the surface $\alpha=0$, we have $\omega_{+}^2 = \omega_B^2$
and $\omega_{-}^2 = \omega_p^2$, but for magnetic fields parallel to the emitting surface one
finds $\omega_{+}^2 = \omega_B^2 + \omega_p^2$ and $\omega_{-}^2 =0$.
Therefore, as we approach $\alpha=\pi/2$, the resonance corresponding to the
plasma frequency in the $\alpha=0$ case is shifted to lower and lower energies.

%%%%%%%%%%%%%%%%%%%%%%%%% FIGURE %%%%%%%%%%%%%%%%%%%%%%%%%%%%%%%%%
\begin{figure}
\resizebox{\hsize}{!}{\includegraphics{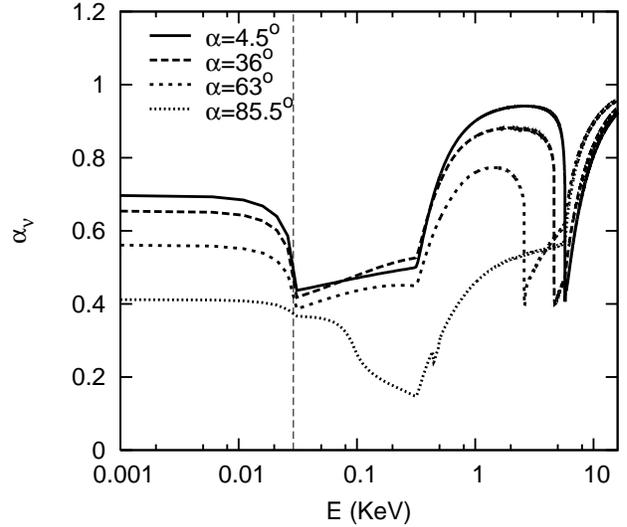}}
\caption{Same as Fig. \ref{spcf13} but
including the effect of the motion of ions. The vertical line corresponds
to the ion cyclotron energy.
}
\label{fig5}
\end{figure}
%%%%%%%%%%%%%%%%%%%%%%%%% FIGURE %%%%%%%%%%%%%%%%%%%%%%%%%%%%%%%%%

The same but for $B=10^{13}$ G is shown in
Fig. \ref{spcf13}. Qualitatively we obtain the same behaviour, but the plasma
frequency is larger and the resonant energy is shifted consistently
to higher values. Notice that the frequency below which the spectrum is depressed
depends weakly on the magnetic field $\omega_p^2/\omega_B \propto B^{1/5}$, and 
for standard values of the magnetic field falls in the range 0.1-0.2 keV. 
In this energy range the effect of interstellar medium absorption can 
make it difficult to distinguish between the two effects.

%%%%%%%%%%%%%%%%%%%%%%%%%%%%%%%%%%%%%%%%%%%%%%%%%%%%%%%%%%%%%%%%%%%%%%%%%%%%
\subsection{The effect of motion of ions.}

When we were finishing this paper, our attention was
drawn to a very recent preprint (\cite{Lai04}) with similar results to ours.
In this work, the authors included terms related to the motion of ions
in the dielectric tensor, which result in smaller reflectivity (larger emissivity) 
at frequencies below the ion cyclotron frequency ($\omega_{B_i} = \frac{ZeB}{m_i c}$). 
The way in which the effect of ions is included is a crude simplification:
as if they were free ions although they are in a lattice. It is quite doubtful
that this approximation can actually represent reality, but it gives some
interesting results that are an indication that more work is needed along that line.
In short, free ions can be introduced in the calculations just by modifying the 
components of the dielectric tensor (\ref{eqn:dielectric}) in the following way
(\cite{Ginz}, \cite{Lai04}):
\begin{eqnarray}
\left( \begin{array}{c} R \\ L \end{array} \right) &=& 1 - 
\frac{\omega_{p}^{2} + \omega_{p_i}^{2}}
{(\omega \mp \omega_{B})(\omega \pm \omega_{B_i}) + i \omega \nu^D_{\perp} A^{\mp}}; \\
P &=& 1- \frac{\omega_{p}^{2} + \omega_{p_i}^{2}}{\omega^{2}+i\omega \nu^D_{\parallel}},
\end{eqnarray}
where $\omega_{p_i}=(4 \pi Z^2 e^2 n_i/m_i)^{1/2}$ is the ion plasma frequency and
\begin{eqnarray}
A^{\mp} = 1 \mp \frac{\omega_{B_i}}{\omega}(1 - Z^{-1}) + \frac{m_e}{m_i}.
\end{eqnarray}
In Fig. \ref{fig5} we show the normalized emissivity integrated over all incident angles,
for $T=10^6$K and $B=10^{13}$ G,  including the effect of ions.
The vertical line indicates the energy corresponding to the ion cyclotron frequency.

In order to understand the effect of ions we follow the same argument after Eq. (\ref{n2app}).
Neglecting damping for simplicity, and considering that the ion plasma frequency is much smaller 
than the electron plasma frequency ($\omega_{p_i}\ll\omega_{p}$), the dispersion relation is simply
\begin{eqnarray}
n^2 = \left( \begin{array}{c} L \\ R \end{array} \right) = 
1-\frac{\omega_{p}^2}{(\omega \pm \omega_{B})(\omega \mp \omega_{B_i})}.
\end{eqnarray}
When $\omega > \omega_{B_i}$ there is not much difference with respect to the case in which 
ions are not included. But when $\omega < \omega_{B_i}$, the new term in the denominator
changes sign and the refractive index of the mode that
acquired a large imaginary part when the ion contribution was not
included becomes
\begin{eqnarray}
n^2 = 1 + \frac{\omega_{p}^2}{ \omega_{B} \omega_{B_i}},
\end{eqnarray}
and the mode is no longer damped. 

%%%%%%%%%%%%%%%%%%%%%%%%%%%%%%%%%%%%%%%%%%%%%%%%%%%%%%%%%%%%%%%%%%%%%%%%
\section{Spectral energy distribution.}

The total spectral emission of the star, the specific luminosity, is obtained 
by integrating the specific intensity over the solid angle pointing outward 
and over the surface of the star,   
\begin{eqnarray}
%L_\nu \int\; I_\nu\; \cos\theta\, dS\, d\Omega
L_\nu=R^2 \int_0^{2\pi}d\phi\int_0^{\pi}\sin\theta\,d\theta
\int_0^{2\pi}d\phi^\prime \\ \nonumber
\int_0^{\pi/2}d\theta^\prime \; 
I_\nu(\theta,\phi,\theta^\prime,\phi^\prime)\; 
\cos\theta^\prime \sin\theta^\prime
\end{eqnarray}
where $\theta$ and $\phi$ are the polar and azimuthal angles of a given point at 
the surface and $\theta^\prime$ and $\phi^\prime$ the polar and azimuthal
angles of the direction defining the solid angle element. 

The presence of the magnetic field, however, makes the emission anisotropic,
and the observed specific flux $F_\nu$ is different from the 
integrated emitted flux obtained from the specific luminosity, 
i.e. $L_\nu/4\pi d^2$ for a star at a distance $d$. To calculate the observed 
specific flux we have to integrate only over the observed hemisphere (OH). Denoting 
by $R$ the radius of the star,
\begin{equation}
F_\nu=\frac{R^2}{d^2}\int_{\rm OH} 
I_\nu(\theta,\phi,\theta^\prime,\phi^\prime)\cos{\theta^\prime}\sin\theta
\,d\theta\, d\phi  
\label{obsflux}
\end{equation}
where the angles $\theta^\prime$ and $\phi^\prime$ specify the angular 
direction from the surface element towards the observer.
Note that in this paper we are neglecting general relativistic effects. 
For example, light bending increases the area of the observed hemisphere, 
depending on the compactness ($M/R$) of the neutron star, as discussed in
Page (\cite{Page95}) or Psaltis, \"Ozel, \& DeDeo (\cite{POD00}).

%%%%%%%%%%%%%%%%%%%%%%%%% FIGURE %%%%%%%%%%%%%%%%%%%%%%%%%%%%%%%%%
\begin{figure}
\resizebox{\hsize}{!}{\includegraphics{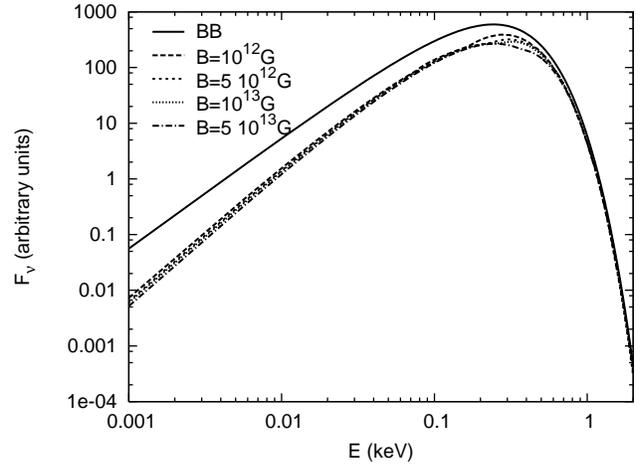}}
\caption{Integrated emitted flux, $L_{\nu}/4\pi D^2$, (in arbitrary units)
for a uniform temperature of $T=10^{6} K$. The magnetic field geometry has been taken to be a 
dipolar distribution and several values of the magnetic field
strength are compared: $B_p= 10^{12}, 5 \times 10^{12}, 10^{13}$, and 
$5 \times 10^{13}$ G (values at the magnetic pole).
}
\label{fluxtot}
\end{figure}
%%%%%%%%%%%%%%%%%%%%%%%%% END FIGURE %%%%%%%%%%%%%%%%%%%%%%%%%%%%%%%%%

In order to get a more accurate feeling of what a realistic emission
would be, in the following we will consider that the magnetic field
has a dipolar geometry, this is
\begin{equation}
B(R,\theta) = \frac{B_p}{2} ( 1 + 3 \cos^2\theta )^{1/2}
\end{equation}
with $B_p$ being the intensity of the magnetic field at the pole.

\begin{figure}
\resizebox{\hsize}{!}{\includegraphics{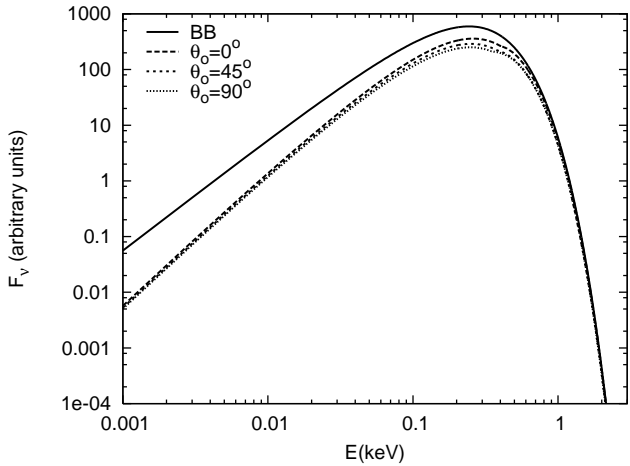}}
\caption{Observed flux, $F_{\nu}$, (in arbitrary units)
for a uniform temperature of $T=10^{6}$ K and three different observation angles. 
The magnetic field geometry has been taken to be a 
dipolar distribution with $B_p=5\times10^{13}$ G.
}
\label{fluxobs}
\end{figure}

The integrated spectrum as a function of the energy is shown in Fig. \ref{fluxtot}
for different intensities of the dipolar magnetic field and a temperature of
$T=10^{6}$ K. For reference, the BB spectrum is also depicted (solid line).
The broadband spectrum is essentially featureless and the flux is systematically
lower than the BB flux for the same temperature; this effect is more significant
at lower energies.
Notice that this flux must be understood as an average over the whole emitting surface, 
which is radiating anisotropically. The real observed flux depends on the particular
location of the observer, as defined by Eq. (\ref{obsflux}). In Fig. \ref{fluxobs}
we show the flux observed from three different directions, forming an angle
of $0^{\circ}$, $45^{\circ}$, and $90^{\circ}$, with the polar axis, respectively. Near the maximum
the differences between observers can be at most a factor 2, but the observer
location becomes less relevant as we move towards either the low and high energy
part of the spectrum. 

At this point, the two main features of the ``metallic surface'' model seem to
be: first, an almost featureless spectrum, and second, an overall flux
smaller than that of a BB at the same temperature, especially at low energies.
However, as we mentioned in the introduction, there is an important point
to consider. The fact that the condensed surface is strongly magnetized makes
the thermal conductivity very different in the directions perpendicular
and parallel to the magnetic field lines.  Similar effects have been pointed out to
be relevant in the envelope (Greenstein \& Hartke \cite{GH83}; Page \cite{Page95})
or in the crust, where a very recent study (\cite{GKP04}), finds
that the anisotropy in the temperature distribution depends very strongly on
the particular geometry of the internal magnetic field, resulting in variations
of temperature of up to a factor 5. In a separate work (\cite{paper2}), we will 
report results from a detailed study of the temperature distribution obtained
from 2D diffusion calculations for different magnetic field geometries.
For the purpose of understanding qualitatively the effects on the observed
spectrum, in this paper we will limit our analysis to the case in which
the temperature distribution has the following angular dependence 
\begin{equation}
T = {T_p} \left[ \cos^2\theta_B + \chi
\sin^2\theta_B \right]^{1/4},
\label{Tdipole}
\end{equation}
where $\theta_B$ is the angle between the field and the
normal to the surface, $\chi$ is the ratio between the thermal conductivities
normal and parallel to the magnetic field, and $T_p$ is the polar temperature
(where $\theta_B=0$). The origin of this distribution has been discussed in 
previous works on neutron star envelopes 
(Greenstein \& Hartke \cite{GH83}; Page \cite{Page95}).
Note that for a dipolar magnetic field $\chi$ is a function of the polar angle
because the magnetic field strength varies with the latitude.

%%%%%%%%%%%%%%%%%%%%%%%%% FIGURE %%%%%%%%%%%%%%%%%%%%%%%%%%%%%%%%%
%%%%%%%%%%%%%%%%%%%%%%%%% END FIGURE %%%%%%%%%%%%%%%%%%%%%%%%%%%%%%%%%

\begin{figure}
\resizebox{\hsize}{!}{\includegraphics{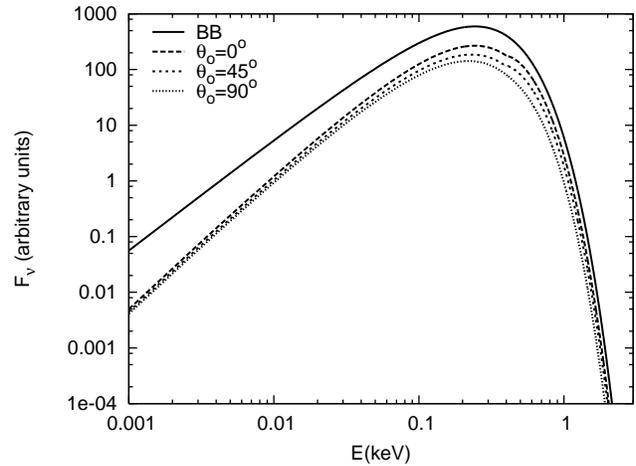}}
\caption{Observed flux, $F_{\nu}$, (in arbitrary units)
for an anisotropic temperature distribution described in Eq.
(\ref{Tdipole}) with $T_{p}=10^{6}$ K and three different observation angles. 
The magnetic field geometry has been taken to be a 
dipole with $B_p= 5\times10^{13}$ G. The corresponding Planckian
spectrum for $T=10^{6}$ K is also shown for comparison.
}
\label{fluxobsani}
\end{figure}

\begin{figure}
\resizebox{\hsize}{!}{\includegraphics{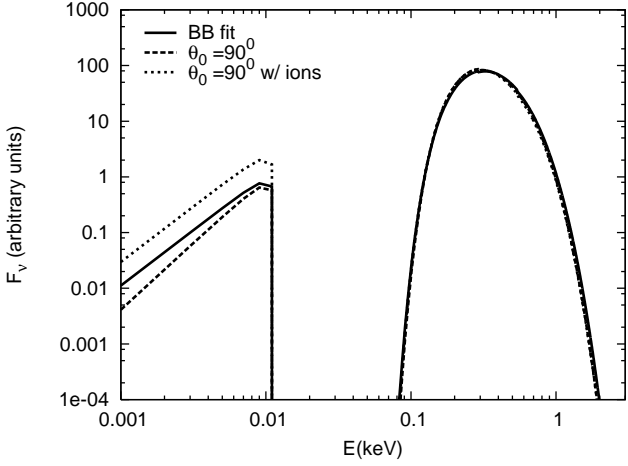}}
\caption{Observed flux, $F_{\nu}$, (in arbitrary units)
for the anisotropic temperature distribution described in Eq.
(\ref{Tdipole}) with $T_{p}=10^{6}$K and $\theta_o = 90^\circ$, as it would be seen
after taking into account interstellar medium absorption
with $n_H=1.4\times10^{20}$cm$^{-2}$. We show results with (dots) and without (dashes)
including the effect of ions.
A uniform temperature, blackbody fit of the X-ray part of the spectrum is also
depicted with solid lines ($T=10^6$K, $n_H=1.3\times10^{20}$cm$^{-2}$) but 
corrected by a factor $1/5$, that is, the apparent estimated value of the $R_\infty/d$
is 2.23 times lower than that of the ``real" model.
}
\label{fig9}
\end{figure}

In Fig. \ref{fluxobsani} we show the same three cases as in Fig. \ref{fluxobs}
but for the anisotropic temperature distribution given by Eq. (\ref{Tdipole}) 
with $T_{p}=10^{6}$ K. The optical band of the spectrum is not very much
altered, but the high energy tail is significantly depressed. This effect,
combined with the low energy depression caused by the high reflectivity
of the metallic surface at low energies, results in a broadband spectrum
that mimics the BB spectrum but with an overall reduced flux by nearly
a factor of 10. This means that, for a fixed distance to the source,
the observed flux from such a particular neutron star surface will look
like a Planckian spectrum, but the apparent area of the source (and therefore the radius) 
would be underestimated by a large factor. To make this point more explicit,
in Fig \ref{fig9} we plot the observed flux (dashes) from a model with $B_p=5\times10^{13}$,
$T_p= 10^{6}$ and $\theta_o = 90^\circ$, as seen
after taking into account interstellar medium absorption
with $n_H=1.4\times10^{20}$ cm$^{-2}$, compared with
a uniform temperature, blackbody model that fits the X-ray part of the spectrum (solid line).
The parameters of the BB model are: $T=10^6$ K, $n_H=1.3\times10^{20}$ 
cm$^{-2}$, and a relative normalization factor ($\propto (R_\infty/d)^2$) of $1/5$. 
Consequently, the apparent estimated value of the $R_\infty/d$
is 2.3 times lower than that of the ``real'' model, despite the X--ray 
spectrum being very similar.

For comparison, we have also included 
the effects of ions in Fig. \ref{fig9} (dotted line).  Notice that the optical flux of this
model is a factor 3 larger than the BB fit prediction,
similarly to what has been observed in isolated neutron stars. Nevertheless,
it must be stressed that this is just a crude approximation, with no pretension
of being the real answer to the observed optical excess, but it serves 
to illustrate how important it is to understand details
about the magnetic field structure, the properties of the solid lattice,
and the temperature distribution, before one is able
to make robust estimates of the neutron star properties (e.g. radius).

%%%%%%%%%%%%%%%%%%%%%%%%% FIGURE %%%%%%%%%%%%%%%%%%%%%%%%%%%%%%%%%

%%%%%%%%%%%%%%%%%%%%%%%%%%%%%%%%%%%%%%%%%%%%%%%%%%%%%%%%%%%%%%%%%%%%%%%%
\section{Final remarks.}

In this paper we have revisited the {\it bare neutron star surface} model
first studied in detail in B80, which in the last years is becoming popular
again and attracting the attention of other groups (\cite{Lai01},
Turolla et al. \cite{TZD04}, \cite{Lai04}).
Our results for constant temperature magnetized surfaces confirm qualitatively 
those reported earlier by other authors, with small quantitative differences 
due to the fact that in previous works (Turolla et al. \cite{TZD04}) some approximations
(neglecting one mode) were made,  which made the results dependent on the cutoff value 
of the imaginary part of the refraction index. In general, models with uniform 
temperature show a broadband spectrum that is very close to Planckian
at energies above $\omega_p^2/\omega_B$, and significantly depressed
(up to a factor 10) in the optical band. The spectrum is almost featureless,
with only some small bumps at energies where the interstellar medium
absorption makes it difficult to distinguis and fine tune between different
parameters. However, in our opinion, there is a key point that is barely
addressed in previous works and needs more attention: in the crust and
the condensed outer layer the assumption of a homogeneous
temperature distribution is inconsistent because magnetic fields of the
order of or larger than $10^{13}$ G imply some degree of anisotropy in 
the thermal conductivity. The analogous effect in the inner crust has 
recently been studied (\cite{GKP04}), finding large variations of temperature
when the magnetic field is confined to the crust. Notice that, if in the inner
regions we have superconducting protons, as seems
to be the case, this situation is very likely.
A detailed analysis of the realistic, self--consistent emission from bare
neutron star surfaces requires, therefore, multidimensional transport
calculations with the presence of magnetic fields, and using appropriate
boundary conditions (accordingly with the calculated $\alpha_{\nu}$).
Such calculations, as well as fits to real data, are in progress (\cite{paper2}) 
and will be reported
elsewhere. In the mean time, one can guess what sort of changes to expect
by looking at the emitted spectrum produced by an ad--hoc temperature
distribution, as we discussed in this paper.
This example was very illustrative of one fact: the observed flux of
such an object is very close to a BB spectrum, but we might be underestimating
the area of the emitter (and therefore its size) by a large factor.
In addition, depending on the strength of the magnetic field, and including the
effects of ions, we could even
obtain an optical flux larger (relative to the BB case) than that in the
X-ray band, which is commonly found in all isolated neutron stars with an
optical counterpart. 

%%%%%%%%%%%%%%%%%%%%%%%%% FIGURE %%%%%%%%%%%%%%%%%%%%%%%%%%%%%%%%%

As stated before, general relativistic effects 
have not been included in this work. A first correction is simply to redshift all 
energies and temperatures. This is of crucial importance if spectral features
are present, but only translates into an overall scale factor if we consider
a BB or a featureless spectrum. A second effect might be more relevant. Light
bending increases the observed emitting area, smearing out partially the
differences between different observers, depending on how compact the object is.
In order to make precise parameter estimates at least these two major
corrections should be included.
In summary, there is much physics to be understood
and analyzed in the near future before drawing robust conclusions on the
nature of isolated compact objects (neutron stars vs. strange stars), and
measuring with precision their radii and masses.

%%%%%%%%%%%%%%%%%%%%%%%%% END FIGURE %%%%%%%%%%%%%%%%%%%%%%%%%%%%%%%%%

%\bigskip
\begin{acknowledgements}
This work has been supported by the Spanish Ministerio
de Ciencia y Tecnolog\'{\i}a grant AYA 2001-3490-C02.
JAP is supported by a {\it Ram\'on y Cajal} contract from the Spanish MCyT.
We thank Fred Walter, Jim Lattimer and Dany Page
for interesting comments and discussions.
\end{acknowledgements}

%%%%%%%%%%%%%%%%%%%%%%%%%%%%%%%%%%%%%%%%%%%%%%%%%%%%%%%%%%%%%%%%%%%%%%%%
\begin{appendix}

\section{Reflected wave amplitudes.}
The system of equations (\ref{systemE}) can be solved for the electric
field of the reflected wave in terms of the incident one. The result reads
as follows:
\begin{eqnarray}
E^{''}_{\perp} =& {\cal D}^{-1}& \left\{ \left[ B_{1}(1-w_{1}) \left(A_{2}\sin i + C_2 \cos i \right) -
\right. \right. \nonumber \\
&& \left. \left. B_{2}(1-w_{2}) \left(A_1 \sin i + C_{1} \cos i \right) \right] \right. E_{\perp}+
\nonumber \\
&& \left. 2B_{1}B_{2} (w_{1}-w_{2}) \sin i \cos i ~ E_{\parallel}\right\}
\nonumber \\
E^{''}_{\parallel} = &{\cal D}^{-1}& \left\{ 2(A_{1}C_{2}-A_{2}C_{1})~E_{\perp}+\right.  \nonumber \\
&+&\left. \left[B_{1}(1+w_{1}) \left( A_{2} \sin i - C_2 \cos i \right) -
\right. \right. \nonumber \\
&& \left. \left. B_{2}(1+w_{2}) \left( A_1 \sin i - C_1 \cos i \right) \right]E_{\parallel} \right\}
\label{eq:campos}
\end{eqnarray}
where 
\begin{eqnarray}
w_{m} &=& \frac{\sqrt{n_{m}^{2}-\sin^{2} i}}{\cos i}~,
\\
{\cal D} &=& B_{1}(1+w_{1}) \left( A_{2} \sin i + C_2 \cos i \right) -
\nonumber \\
&& B_{2}(1+w_{2}) \left( A_1 \sin i + C_1 \cos i \right)
\end{eqnarray}
From these equations the original expressions in B80 can be recovered by 
using the relation
\begin{equation}
C_m = \sin^2 i + A_m~ w_m \sin i \cos i.
\end{equation}
We have found a typo in the first line of Eq. (19) in B80, where the last ${\cal A}_2$
should be a ${\cal B}_2$.
Notice that Turolla et al. (\cite{TZD04}) expressed the amplitudes of the reflected
waves in a different way. The reason for this apparent difference is the same that led to different
expressions for the coefficient $a_m$ defined by Eq. (\ref{eq:amplx}): they
used a different linear combination of the three equations
arising from Eq. (\ref{max1}). We have checked numerically that both formulations
are equivalent when $n_m$ is a root of Eq. (\ref{eq:cuartica}).

\end{appendix}

%%%%%%%%%%%%%%%%%%%%%%%%%%%%%%%%%%%%%%%%%%%%%%%%%%%%%%%%%%%%%%%%%%%%%%%%

\end{document}